\documentclass[aps,prb,twocolumn,superscriptaddress]{revtex4-2}

\usepackage{graphicx}
\usepackage{amsmath}
\usepackage{mathtools,amssymb,lipsum}
\usepackage{titlesec}
\usepackage{braket}
\usepackage{soul}
\usepackage{comment}
\usepackage{xcolor}
\usepackage{scalerel}

\begin{document}
	
	
	\title{Impact of measurement backaction on nuclear spin qubits in silicon}
	
	
	\author{S. Monir*}
	\affiliation{School of Physics, The University of New South Wales, Sydney, NSW 2052, Australia}
	\affiliation{Silicon Quantum Computing Pty Ltd., Level 2, Newton Building, UNSW Sydney, Kensington, NSW 2052, Australia}
	
	\author{E. N. Osika}
	\affiliation{School of Physics, The University of New South Wales, Sydney, NSW 2052, Australia}
	\affiliation{Silicon Quantum Computing Pty Ltd., Level 2, Newton Building, UNSW Sydney, Kensington, NSW 2052, Australia}
	
	\author{S. K. Gorman}
	\affiliation{Silicon Quantum Computing Pty Ltd., Level 2, Newton Building, UNSW Sydney, Kensington, NSW 2052, Australia}
	\affiliation{Centre for Quantum Computation and Communication Technology, School of Physics, University of New South Wales, Sydney, NSW 2052, Australia}
	
	\author{I. Thorvaldson}
	\affiliation{Silicon Quantum Computing Pty Ltd., Level 2, Newton Building, UNSW Sydney, Kensington, NSW 2052, Australia}
	\affiliation{Centre for Quantum Computation and Communication Technology, School of Physics, University of New South Wales, Sydney, NSW 2052, Australia}

	\author{Y.-L. Hsueh}
	\affiliation{School of Physics, The University of New South Wales, Sydney, NSW 2052, Australia}
	\affiliation{Silicon Quantum Computing Pty Ltd., Level 2, Newton Building, UNSW Sydney, Kensington, NSW 2052, Australia}
	
	\author{P. Macha}
	\affiliation{Silicon Quantum Computing Pty Ltd., Level 2, Newton Building, UNSW Sydney, Kensington, NSW 2052, Australia}
	\affiliation{Centre for Quantum Computation and Communication Technology, School of Physics, University of New South Wales, Sydney, NSW 2052, Australia}
	
	\author{L. Kranz}
	\affiliation{Silicon Quantum Computing Pty Ltd., Level 2, Newton Building, UNSW Sydney, Kensington, NSW 2052, Australia}
	\affiliation{Centre for Quantum Computation and Communication Technology, School of Physics, University of New South Wales, Sydney, NSW 2052, Australia}
	
	\author{J. Reiner}
	\affiliation{Silicon Quantum Computing Pty Ltd., Level 2, Newton Building, UNSW Sydney, Kensington, NSW 2052, Australia}
	\affiliation{Centre for Quantum Computation and Communication Technology, School of Physics, University of New South Wales, Sydney, NSW 2052, Australia}
	
	\author{M.Y. Simmons}
	\affiliation{Silicon Quantum Computing Pty Ltd., Level 2, Newton Building, UNSW Sydney, Kensington, NSW 2052, Australia}
	\affiliation{Centre for Quantum Computation and Communication Technology, School of Physics, University of New South Wales, Sydney, NSW 2052, Australia}
	
	\author{R. Rahman}
	\affiliation{School of Physics, The University of New South Wales, Sydney, NSW 2052, Australia}
	\affiliation{Silicon Quantum Computing Pty Ltd., Level 2, Newton Building, UNSW Sydney, Kensington, NSW 2052, Australia}
	
	
\begin{abstract}
Phosphorus donor nuclear spins in silicon couple weakly to the environment making them promising candidates for high-fidelity qubits. The state of a donor nuclear spin qubit can be manipulated and read out using its hyperfine interaction with the electron confined by the donor potential. Here we use a master equation-based approach to investigate how the backaction from this electron-mediated measurement affects the lifetimes of single and multi-donor qubits. We analyze this process as a function of electric and magnetic fields, and hyperfine interaction strength. Apart from single nuclear spin flips, we identify an additional measurement-related mechanism, the nuclear spin flip-flop, which is specific to multi-donor qubits. Although this flip-flop mechanism reduces qubit lifetimes, we show that it can be effectively suppressed by the hyperfine Stark shift. We show that using atomic precision donor placement and engineered Stark shift, we can minimize the measurement backaction in multi-donor qubits, achieving larger nuclear spin lifetimes than single donor qubits.
\end{abstract}
	
	
\maketitle

\section{Introduction}

Phosphorus (P) nuclear spins in silicon (Si) are a promising candidate for a fault-tolerant quantum computing architecture due to their weak coupling to the environment ~\cite{Kane1998, Hill2015, OGorman2016, Pica2016}. Such a system has demonstrated seconds long coherence times in isotopically purified $^{28}$Si \cite{Muhonen2014}. The spin $\frac{1}{2}$ nucleus of the P donor atom couples to its bound electronic spin through the hyperfine interaction which can be tuned by an applied electric field. By combining electron spin resonance with single-shot electron spin readout we can measure the state of the nuclear spin qubit through the hyperfine interaction \cite{Pla2013, madzik2021precision}. During such a readout process the electron, under appropriate applied voltages, tunnels between the qubit and a nearby reservoir such as a Single Electron Transistor (SET) island, used as a charge sensor \cite{Elzerman2004, Morello2010, Keith2019}. From the perspective of the nuclear spin, the hyperfine interaction appears as being switched on (when the electron is on the qubit) and off (when the electron is on the SET). The mixing of the nuclear spin states due to these instantaneous, non-adiabatic changes in the hyperfine interaction during electron tunneling events affects the stability of the nuclear spin. In previous works \cite{Pla2013}, this measurement backaction has been observed to reduce the lifetimes of single nuclear spin states of P donors prepared in the $\ket{\Uparrow}$ spin state.

There has been much recent progress in implementing few qubit quantum processors with multi-donor systems, where electrons shared between the P atoms can couple to multiple nuclear spins \cite{madzik2021precision, He2019_SWAP, Kranz2022}. Such multi-donor quantum dot qubits also offer advantages in terms of longer spin relaxation times \cite{Hsueh_2014}, highly tunable inter-qubit exchange coupling \cite{Wang_exch_2016}, and improved addressibility with high operation speeds due to larger ranges of hyperfine values \cite{Buch2013, Hile2018}. How the measurement backaction impacts such multi-nuclear systems and how they can be controlled by external means have remained open questions. 
	
In this article we study the effect of the tunneling electron on the lifetimes of nuclear spin states in single and multi-donor qubits. Using a master equation based approach we investigate the influence of hyperfine interaction strength, magnetic fields and electric fields on nuclear spin transition probabilities for both single and multi-donor qubits. We identify an additional flip-flop mechanism between the nuclear spins resulting from measurement backaction in multi-donor systems. This effect, along with the typically larger values of the hyperfine coupling, can reduce the nuclear spin lifetimes in multi-donor quantum dots. However, as we show, these flip-flop transitions can be suppressed when the differences between the hyperfine couplings to various nuclear spins are large. Also, the total hyperfine coupling of multi-donor qubit is strongly dependent on donor positions, thus it can be controlled with donor placement. In this paper we provide exact instructions on how to design multi-donor qubits to minimize the effect of measurement backaction even below that of single-donor qubits.
	
\begin{figure}[]
	\centering
	\includegraphics[width = 7.5 cm]{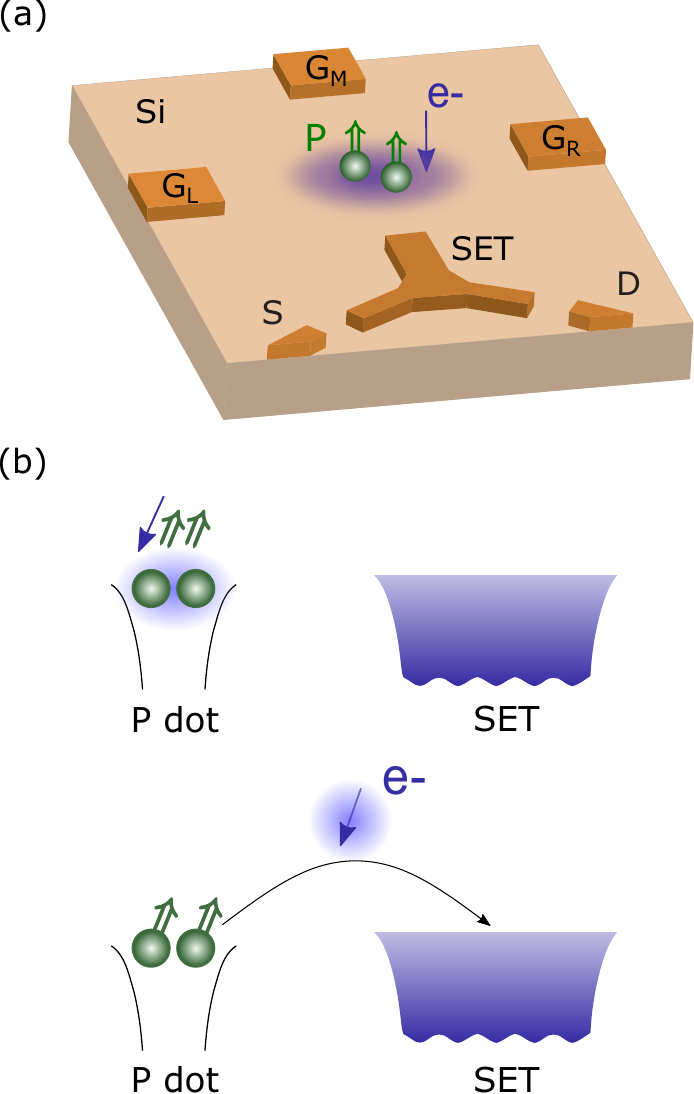}%
	\caption{\textbf{Schematic of the multi-donor qubit and spin readout sensor}. a) P donors can be precision placed with $\pm 0.385$~nm accuracy in Si crystal and surrounded by in-plane P doped gate electrodes (G$_\mathrm{L}$, G$_\mathrm{M}$ and G$_\mathrm{R}$), source (S), drain (D) and SET charge sensor by STM lithography \cite{Fuechsle2012}. The donor potential traps electron which can tunnel to and from the SET under appropriate in-plane gate biases. b) Effect of measurement by an SET sensor on the donor nuclear spin states. The electron non-adiabatically tunneling from the donor to the SET leaves the nuclear spins no longer in their eigenstate resulting in a finite probability of a nuclear spin flip.
    \label{intro}}
\end{figure}

\section{Method}

The system we investigate in this work is shown schematically in Fig.~\ref{intro}(a). It consists of a single- or multi-donor qubit with an electron confined by the donor potential and the surrounding gates and SET used for qubit manipulation and measurement. The measurement process involves an electron tunneling between the dot and the SET -- see Fig.~\ref{intro}(b). As a result of this non-adiabatic tunneling process, the overall qubit system can be brought to a state that is not one of its eigenstates. The spins evolve over time with a mixture of different eigenfrequencies, introducing a finite probability of nuclear spin flips. 


We simulate the dynamics of the spin system under qubit control pulses using the master equation in Lindblad form. The combined donor-electron system comprises of the donor nuclear spin levels ($\ket{\Uparrow}$ and $\ket{\Downarrow}$) and electronic levels ($\ket{\uparrow}$, $\ket{\downarrow}$ and off the dot (on the SET)). Because of the continuum of electronic levels in the SET, we do not distinguish between $\ket{\uparrow}$ and $\ket{\downarrow}$ spin states when the electron is located in the SET. The density operator $\rho$ of this open quantum system evolves according to the master equation in Lindblad form ($\hbar = 1$),
\begin{equation} \label{master_eqn}
	\partial_t \rho = -i[H, \rho]_{-}+\sum_{\mu} \left( L_{\mu} \rho L_{\mu}^{\dagger}-\frac{1}{2}[L_{\mu}^{\dagger}L_{\mu}, \rho]_{+}\right)
\end{equation}
In this equation $H$ is the Hamiltonian of the system operating on a Hilbert space $\mathcal{H}$ that can be decomposed as $\mathcal{H}_n^{(1)} \otimes \cdots \mathcal{H}_n^{(j)} \otimes \cdots \otimes \mathcal{H}_n^{(m)} \otimes \mathcal{H}_e$ where, $\mathcal{H}_n^{(j)} (\mathcal{H}_e)$ is the subspace of the $j^{th}$ donor nuclear spin (electron). The $L_{\mu}$'s are Lindblad operators corresponding to electron tunnelings and relaxation pathways. 
	
We consider two cases depending on the location of the electron. 1) When the electron is on the donor dot, the Hamiltonian is written as,
\begin{equation}
	H_{donor} = \sum_{j = 1}^{m} \gamma_n \mathbf{I}_j \cdot \mathbf{B} + \gamma_e \mathbf{S} \cdot \mathbf{B}+\sum_{j = 1}^{m} \mathbf{I}_j \cdot \mathbf{A}_j \cdot \mathbf{S}
\end{equation}
where, $\gamma_n(\gamma_e)$ is the nuclear(electron) gyromagnetic ratio and $\mathbf{I}$($\mathbf{S}$) is the nuclear(electron) spin operator. We take the electron gyromagnetic ratio of electron $\gamma_e = 27.958$ GHz/T and the P donor nuclear spin gyromagnetic ratio $\gamma_n = -17.217$ MHz/T. The first two terms in the equation are the nuclear and electron Zeeman interactions in an external magnetic field $\mathbf{B}$, respectively and the last term is the hyperfine interaction term where $\mathbf{A}_j$ is the hyperfine tensor between the $j^{th}$ nucleus and the electron. For our simulations we use only the scalar contact hyperfine term $A_j$ since it is the only dominating term for the measurement backaction mechanism (see Supplementary Material II). 2) When the electron is on the SET, we only consider the nuclear Zeeman interaction,
\begin{equation}
	H_{zn} = \sum_{j = 1}^{m} \gamma_n \mathbf{I}_j \cdot \mathbf{B}
\end{equation}
The operator $H_{donor}$  acts on a 2D electronic subspace (spanned by electron $\ket{\uparrow}$ and $\ket{\downarrow}$ states) and $H_{zn}$ acts on a 1D electronic subspace (spanned by $\ket{SET}$, i.e. when the electron is on the SET). Let us define two projection operators,
\begin{align}
	\mathbb{P}_{donor} & = \mathbb{I}_{2^m \times 2^m} \otimes \begin{pmatrix}
	1 & 0 \\
	0 & 1 \\
	0 & 0 \\
	\end{pmatrix} \\
	\mathbb{P}_{SET} & = \mathbb{I}_{2^m \times 2^m} \otimes \begin{pmatrix}
	0 \\
	0 \\
	1 \\
	\end{pmatrix}
\end{align}
	
Here, $\mathbb{I}_{2^m \times 2^m}$ is an identity operator acting on the subspace of $m$ nuclear spins. The operator $\mathbb{P}_{donor}$ ($\mathbb{P}_{SET}$) projects any state in the 2D (1D) electronic subspace to the space where the electron is a three-level system spanned by $\{\ket{\uparrow}, \ket{\downarrow}, \ket{SET}\}$. The Hamiltonian $H$, therefore, can be written as,
\begin{equation}
	H = \mathbb{P}_{donor} H_{donor} \mathbb{P}_{donor}^\dagger + \mathbb{P}_{SET} H_{zn} \mathbb{P}_{SET}^\dagger
\end{equation}
	
For the simulations of time evolution, we switch to the eigenbasis and construct the Lindblad operators ($L_\mu$) in that basis. The Hamiltonian and Lindblad operators construction for a single donor has been described in Supplementary Material I. For simplicity, here we consider the system in a noiseless environment. Hence the non-unitary time evolution is caused only by the tunneling electron. The number of $L_\mu$'s can vary depending on the number of possible ways for the electron to tunnel during a particular spin-control pulse. The method can be generalized for an arbitrary number of donors and expanded to include the effects of other relaxation mechanisms such as hyperfine mediated relaxation \cite{Hsueh2023, Camenzind2018, Pines1957} and magnetic noise.

\section{Results}

\subsection{Single nuclear spin flip}

\begin{figure*}[t]
	\centering
	\includegraphics[width = 18 cm]{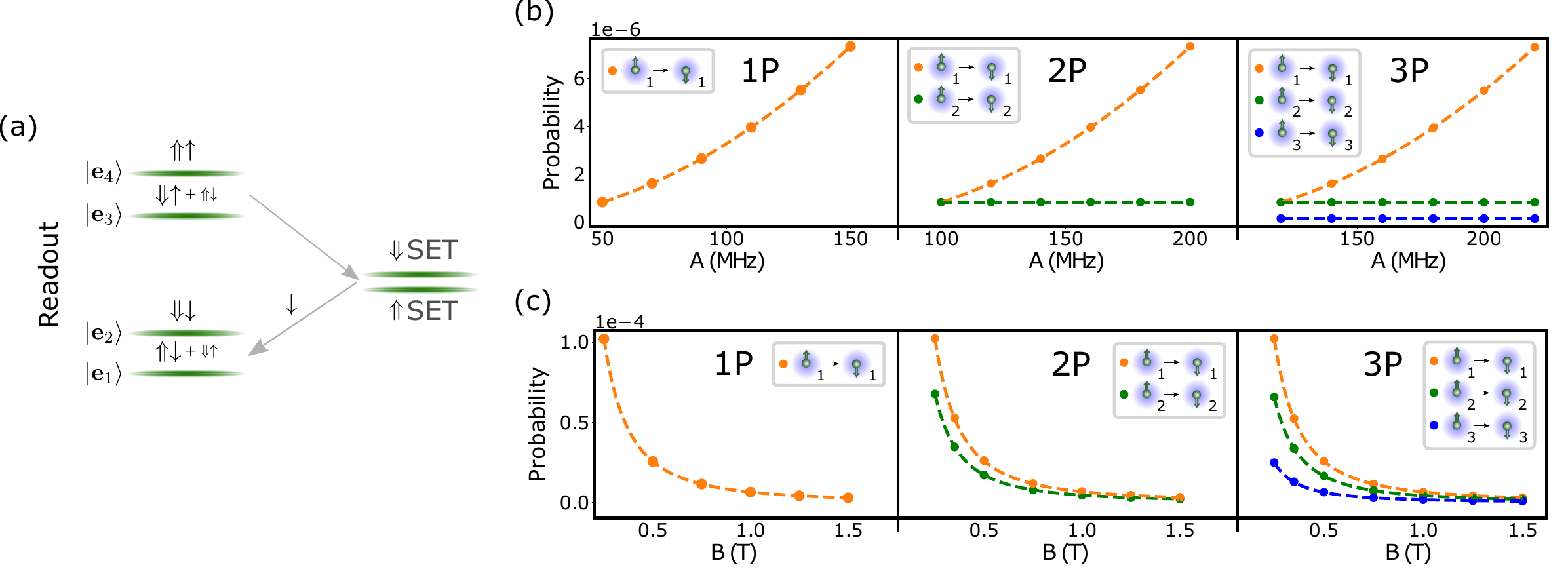}%
	\caption{\textbf{Nuclear spin flip probability in P donor qubits due to the readout pulse}. a) Electron tunneling in a 1P1e qubit during the readout pulse. The labels $\ket{\mathbf{e}_1} - \ket{\mathbf{e}_4}$ represent eigenstates of the 1P1e system. The arrows show the nuclear-electron states forming the eigenstates, with the smaller arrows (i.e $\Downarrow\uparrow$ in $\ket{\mathbf{e}_1}$ and $\Uparrow\downarrow$ in $\ket{\mathbf{e}_3}$) representing typically very small admixtures  of nuclear-electron spin configurations introduced by the hyperfine interaction. b) Dependence of the nuclear spin  flip probability on the total hyperfine interaction A of 1P1e (left panel), 2P1e (middle) and 3P1e (right) qubits at a magnetic field of 1.4 T. For each of these three cases only the hyperfine constant of one donor ($A_1$) is varied. The other hyperfines are fixed at $A_2 = 50$ MHz and $A_3 = 20$ MHz. c) Dependence of the nuclear spin  flip probability on the magnetic field B for 1P1e (left panel), 2P1e (middle) and 3P1e (right) qubits. The hyperfines are fixed at  $A_1 = 100$ MHz, $A_2 = 50$ MHz and $A_3 = 50$ MHz. For the sub-figures (b-g), the dotted lines represent the analytical form of Eq.~(\ref{AB_dependence}).
    \label{fig_AB}}
\end{figure*}

We have verified the method presented in this paper by analysing the single-donor experimental data from Ref.~\cite{plathesis} -- we describe those simulations in detail in Supplementary Material I and II. We also show in Supplementary Material II that the readout causes predominantly one-directional nuclear spin flips, i.e. $\Uparrow \rightarrow \Downarrow$, thus we focus on this transition further in the main text. Additionally, the $\Downarrow \rightarrow \Uparrow$ transition is dominated by the hyperfine mediated relaxation which is a few orders of magnitude stronger than the backaction effects from readout \cite{plathesis}. For a single donor qubit we derive the expression for the $\Uparrow\rightarrow\Downarrow$ transition probability due to readout analytically. We start in the $\ket{\mathbf{e}_4}=\ket{\Uparrow \uparrow}$ state -- see Fig.~\ref{fig_AB}(a) for the scheme of the readout process and the definition of the $\ket{\mathbf{e}_1}-\ket{\mathbf{e}_4}$ eigenstates. After the first tunneling event -- electron tunneling from the donor to the SET -- the system is still in its eigenstate, specifically in $\ket{\Uparrow SET}$, an eigenstate of $H_{zn}$. However, after the second tunneling event -- another electron $\ket{\downarrow}$ tunneling instantaneously from the SET to the donor -- the system is now in the $\ket{\Uparrow \downarrow}$ state which evolves as a mixture of the eigenstates of $H_{donor}$ as,
\begin{align}
	\ket{\Psi (t)} & = \cos\frac{\theta}{2} \ket{\mathbf{e}_1}e^{-i\omega_1 t}-\sin\frac{\theta}{2} \ket{\mathbf{e}_3}e^{-i\omega_3 t} \nonumber \\
	& = \cos\frac{\theta}{2} (\cos\frac{\theta}{2} \ket{\Uparrow \downarrow}+\sin\frac{\theta}{2} \ket{\Downarrow \uparrow})e^{-i\omega_1 t} \nonumber \\
	& -\sin\frac{\theta}{2} (-\sin\frac{\theta}{2} \ket{\Uparrow \downarrow}+\cos\frac{\theta}{2} \ket{\Downarrow \uparrow})e^{-i\omega_3 t}
\end{align}
where, $\omega_1 = -\frac{A}{4}+\frac{1}{2}\sqrt{(\omega_n-\omega_e)^2+A^2}$, $\omega_3 = -\frac{A}{4}-\frac{1}{2}\sqrt{(\omega_n-\omega_e)^2+A^2}$, and $\theta = \tan^{-1}\frac{A}{\omega_n-\omega_e}$. This mixing of pure states by the hyperfine interaction introduces a finite probability for the state of the system to transition from $\ket{\Uparrow \downarrow}$ to $\ket{\Downarrow \uparrow}$. The probability of $\Uparrow \rightarrow \Downarrow$ transition is obtained by time-averaging as follows.
\begin{align} \label{AB_dependence}
	P_{\Uparrow\rightarrow\Downarrow} & = \left< \left|\sin\frac{\theta}{2}\cos\frac{\theta}{2}(e^{-i\omega_1 t}-e^{-i\omega_3 t})\right|^2 \right> \nonumber \\
	 & = \frac{1}{2}\frac{A^2}{A^2+(\omega_n-\omega_e)^2} \nonumber\\
  & \approx \frac{1}{2}\frac{A^2}{((\gamma_n B)-(\gamma_e B))^2} 
	 \propto \frac{A^2}{B^2} 
\end{align}
	
Here we assume that the hyperfine interaction is much smaller than the electron spin Zeeman energy splitting, $A \ll \omega_e$, true for the most recent experiments \cite{Pla2013, Buch2013, Broome2018, Hile2018}. Fig.~\ref{fig_AB} (b,c), left-most panels, shows the comparison between the analytical formula from Eq.~(\ref{AB_dependence}) and the $P_{\Uparrow\rightarrow\Downarrow}$ from master equation simulations for a single donor occupied by a single electron, i.e. 1P1e qubit. The $A^2/B^2$ trend is expected due to the mixing of nuclear-electron spins through the hyperfine interaction and has been discussed in optical detection of $^{31}$P qubits \cite{Fu2004} and quantum dot qubits in Si \cite{Zhao2019}. Since we specifically consider the readout event for an initial state $\ket{\Uparrow\uparrow}$, we obtain a factor of $1/2$ in Eq.~(\ref{AB_dependence}) as opposed to $1/16$ in previous works \cite{Fu2004, Zhao2019}. The transition probabilities are of the order of $10^{-6}$ since the electron spin Zeeman splitting is typically 2-3 orders of magnitude larger than the hyperfine coupling in experiments. 

	
\begin{figure}[t]
\centering
\includegraphics[width = 7.5 cm]{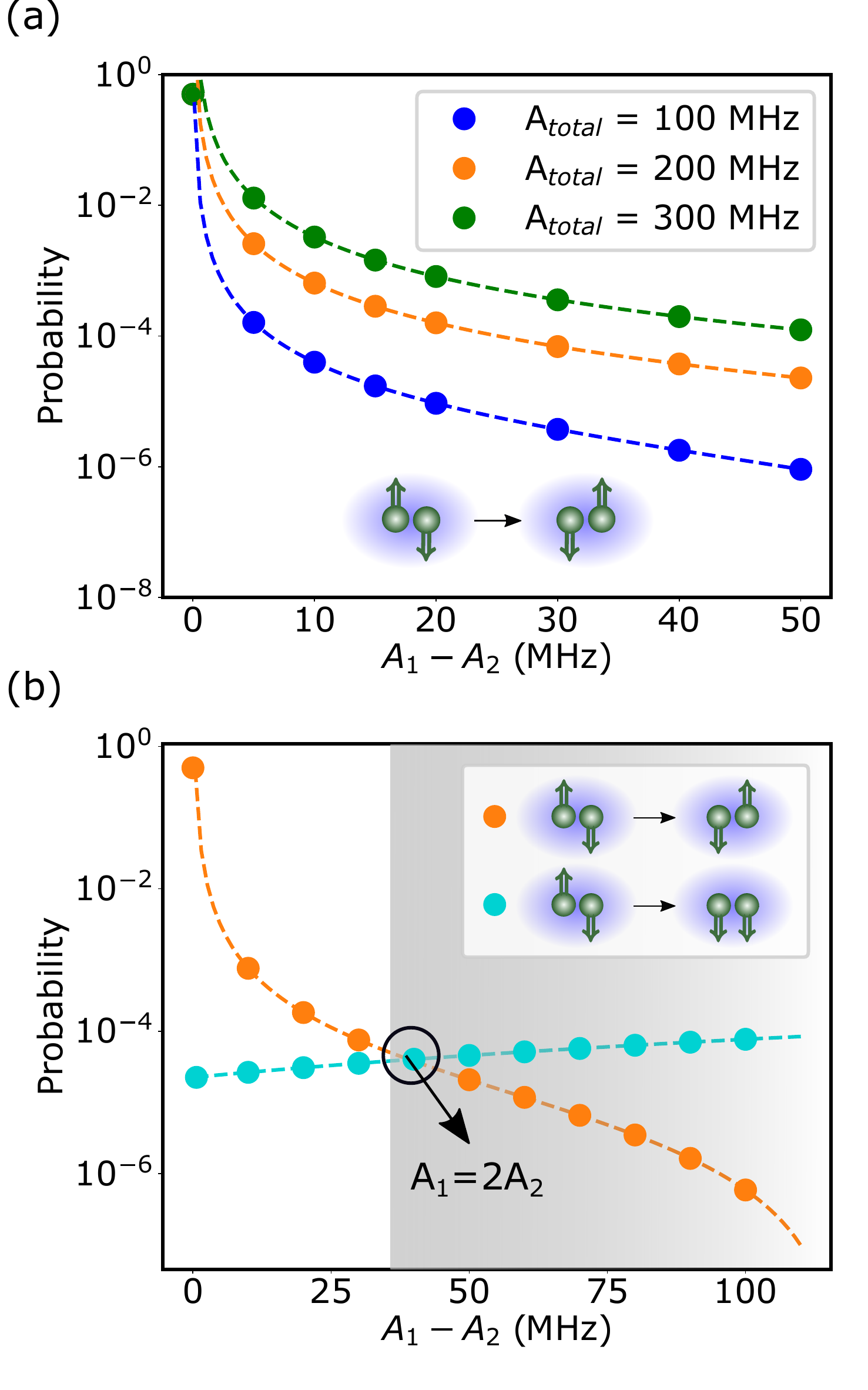}%
\caption{\textbf{Nuclear spin flip-flop probabilities in multi-donor dots due to the readout pulse}. a) Variation of nuclear spin flip-flop probability with $A_1-A_2$ in a 2P1e qubit for total hyperfine values of $100,~200$ and $300$ MHz. The dotted lines represent the formula in Eq.~ (\ref{AB_dependence}) b) Comparison between nuclear spin flip ($\Uparrow\Downarrow\rightarrow\Downarrow\Downarrow$) and flip-flop ($\Uparrow\Downarrow\rightarrow\Downarrow\Uparrow$) transition probabilities for a 2P1e qubit with total hyperfine $A_{total}  = 117$ MHz as a function of $A_1-A_2$. The orange and cyan dotted lines represent formulas of Eq.~(\ref{AB_dependence}) and Eq.~(\ref{flip-flop}), respectively. The shaded region represents the Stark shift values for which the total transition probability of the first nuclear spin (flip+flip-flop) is less than the transition probability of a 1P1e qubit. The transition probability of second nuclear spin is even smaller in this region since it has a smaller hyperfine constant than the first nuclear spin. Therefore, in this region the 2P1e qubit can experience less effects from measurement backaction than a 1P1e qubit.
\label{2P_3P}}
\end{figure}
    
For multi-donor qubits we can also show that the $\Uparrow\rightarrow\Downarrow$ transitions due to measurement backaction vary as $A^2/B^2$. 
Moreover, the $\Uparrow\rightarrow\Downarrow$ transition probability of a particular nuclear spin does not depend on the hyperfine value or the state of the other nuclear spins in the dot ($\Gamma_{\Uparrow \Uparrow \rightarrow \Uparrow \Downarrow} = \Gamma_{\Downarrow \Uparrow \rightarrow \Downarrow \Downarrow}, \Gamma_{\Uparrow \Uparrow \rightarrow \Downarrow \Uparrow} = \Gamma_{\Uparrow \Downarrow \rightarrow \Downarrow \Downarrow}$). That is because the mixing of $\ket{\Uparrow}$ and $\ket{\Downarrow}$ states of a particular nuclear spin only depends on its own hyperfine interaction strength with the electron.  
In Fig.~\ref{fig_AB}(b) we consider 2P1e (middle panel) and 3P1e (right panel) systems, i.e. qubits consisting of 2 and 3 P donors respectively, and plot the nuclear spin flip probabilities as a function of total hyperfine interaction, i.e. the sum of all the donor A constants in a given dot. For each of these systems only one hyperfine constant ($A_1$) is varied. The resultant nuclear spin flip probability of the nucleus with modulated hyperfine shows a $A^2$ dependence. We can see that the flip probabilities of other nuclear spins remain constant, even though the total hyperfine of the multi-donor qubit is changing. This highlights the independent behaviour of each nuclear spin with respect to its own hyperfine interaction with the electron spin. In Fig.~\ref{fig_AB}(c) the magnetic field is varied and in both 2P1e and 3P1e cases each nuclear spin flip probability shows a $1/B^2$ dependence, similar to the 1P case. Although not explicitly mentioned in the figures, all the dashed lines plotted in Fig.~\ref{fig_AB}(b, c) follow $\frac{1}{2} A_i^2/(\omega_n - \omega_e)^2$ relation from Eq.~(\ref{AB_dependence}), where $A_i$ is the hyperfine constant of each individual nuclear spin.

\subsection{Nuclear-nuclear spin flip-flop}

For a multi-donor qubit, in addition to the $\ket{\Uparrow \downarrow}$ and $\ket{\Downarrow \uparrow}$ mixing of a single nuclear spin with the electron spin, there is also coupling of the $\ket{\Uparrow\Downarrow}$ and $\ket{\Downarrow\Uparrow}$ states due to the hyperfine interaction. This is an interaction between the nuclear spins mediated by the electron spin and can result in nuclear spin flip-flop transitions. Hyperfine-mediated nuclear-nuclear spin interaction has previously been studied in quantum dots and it plays a major role in dictating the nuclear spin dynamics in the presence of electrons \cite{Latta2011, Maletinsky2007, Deng2005, deSousa2003}. This is different from the direct nuclear spin dipole-dipole interaction, which we discuss separately in Supplementary Material III. During electron tunneling events this effective interaction switches on and off and results in an extra nuclear spin transition pathway in multi-donor qubits.
 
We can obtain an analytical formula for the measurement-driven nuclear-nuclear spin flip-flop transition probability by considering an effective low-energy spin Hamiltonian for the system. The full spin Hamiltonian of a 2P1e system is written as,
\begin{align}
	    H = \omega_n \left(I_{1z} + I_{2z} \right) + \omega_e S_z + A_1 \mathbf{I}_1 \cdot \mathbf{S} + A_2 \mathbf{I}_2 \cdot \mathbf{S}
\end{align}
At the typical experimental magnetic fields of qubit operation ($B \backsim 1.4$ T), the electron Zeeman is the dominant interaction and we can consider $H_0 = \omega_n\left(I_{1z} + I_{2z} \right) + \omega_e S_z$ as the unperturbed Hamiltonian and $A_1 \mathbf{I}_1 \cdot \mathbf{S} + A_2 \mathbf{I}_2 \cdot \mathbf{S}$ as the perturbation. The eigenspectra of $H_0$ has two subspaces corresponding to the electron $\ket{\uparrow}$ and $\ket{\downarrow}$ states. Even after the addition of the perturbation (hyperfine terms), the subspaces of electron spin corresponding to majority $\ket{\uparrow}$ and $\ket{\downarrow}$ are well separated in energy and therefore, we can perform a Schrieffer-Wolff transformation to obtain a $4\times4$ effective Hamiltonian for the low-energy subspace (majority electron $\ket{\downarrow}$) \cite{Bravyi2011, Latta2011}. We perform the perturbative expansion up to second order which gives less than a kHz deviation in eigenvalues with the solution of the full Hamiltonian. The effective Hamiltonian is in a block-diagonal form and in the subspace corresponding to $\{\ket{\Uparrow\Downarrow}, \ket{\Downarrow\Uparrow}\}$ it can be written as,
\begin{align}
	    H_{eff} & = \begin{pmatrix}
	    -\frac{A_1^2}{4(\omega_e-\omega_n)}-\frac{A_1}{4}+\frac{A_2}{4} & -\frac{A_1 A_2}{4(\omega_e-\omega_n)} \\
	    -\frac{A_1 A_2}{4(\omega_e-\omega_n)} & -\frac{A_2^2}{4(\omega_e-\omega_n)}-\frac{A_2}{4}+\frac{A_1}{4}
	    \end{pmatrix} \nonumber \\
	    & = \begin{pmatrix}
	    \frac{-\Delta}{2} & \frac{\tau}{2} \\
	    \frac{\tau}{2} & \frac{\Delta}{2}
	    \end{pmatrix}
\end{align}
where, $\Delta = \frac{A_1-A_2}{2}(1+\frac{A_1+A_2}{4(\omega_e-\omega_n)}) \approx \frac{A_1-A_2}{2}$ and $\tau=-\frac{A_1 A_2}{2(\omega_e-\omega_n)} \approx -\frac{A_1 A_2}{2\omega_e}$. We have chosen the electron $\ket{\downarrow}$ spin Zeeman energy as our energy origin. The off-diagonal elements in $H_{eff}$ are due to the second order processes such as $\Downarrow\Uparrow\downarrow \rightarrow \Downarrow\Downarrow\uparrow \rightarrow \Uparrow\Downarrow\downarrow$. Each time the electron $\ket{\downarrow}$ tunnels in to the qubit, the nuclear spins flip-flop because of this second order interaction. This results in the following average flip-flop probability.
\begin{align}
	    P_{\Uparrow\Downarrow \rightarrow \Downarrow\Uparrow} & = \left<\frac{\tau^2}{\tau^2+\Delta^2} \sin^2 \left( \sqrt{\tau^2+\Delta^2}\frac{t}{2}\right)\right> \nonumber \\
	    & \approx \frac{1}{2} \frac{(\frac{A_1 A_2}{2\omega_e})^2}{(\frac{A_1 A_2}{2\omega_e})^2+(\frac{A_1-A_2}{2})^2}
\end{align}
For the hyperfine values difference of the order of 10 MHz range typically found in experiments,  $|A_1-A_2| >> \frac{A_1 A_2}{\omega_e}$. Therefore, the flip-flop probability depends on the hyperfine values as,
\begin{equation} \label{flip-flop}
	    P_{\Uparrow\Downarrow \rightarrow \Downarrow\Uparrow}  \approx \frac{1}{2} \frac{(\frac{A_1 A_2}{2\omega_e})^2}{(\frac{A_1-A_2}{2})^2}
	     \propto \left(\frac{A_1 A_2}{A_1-A_2}\right)^2
\end{equation}
Here we can see that the flip-flop probability depends on the difference of the two hyperfines values. If the difference in the hyperfine values is small, the electron is almost equally populated on the two donors and the configurations $\ket{\Uparrow\Downarrow}$ and $\ket{\Downarrow\Uparrow}$ are close in energy. In this situation, the mediated interaction is strong which in turn increases the flip-flop rate. The $|A_1-A_2|$ term can, however, be intentionally enhanced in real devices. In a 2P donor qubit the difference in hyperfine values can be engineered by applying an external electric field (Stark shift) while for 3P and above, even without any electric field applied, the hyperfine constants are naturally different due to the asymmetric donor arrangement within the Si crystal structure (except from some rare symmetric cases).

Fig.~\ref{2P_3P}(a) shows the dependence of the flip-flop probability on the Stark shift ($A_1-A_2$) for a 2P1e system during the readout pulse for total hyperfine values of 100, 200 and 300 MHz. Dots represent master equation calculations and dashed lines represent the formula of Eq.~(\ref{flip-flop}). Here we can see that the nuclear spin flip-flop probability indeed decreases with the square of the Stark shift. We can also see that for a given Stark shift the flip-flop rate is larger for a larger total hyperfine, i.e. when the donors are closer together. Larger hyperfine values mean that the electron is coupled more strongly to the nuclei and the mediated interaction is stronger.

In terms of the effects from measurement backaction, a single donor might seem like an intuitively best choice to achieve high-fidelity qubit due to the absence of the nuclear spin flip-flop transitions. However, we will show that specially designed multi-donor qubits can in fact demonstrate superior fidelity. The hyperfine constant of a single P donor in Si is equal approximately to 117 MHz \cite{Feher1959}, which varies very little with electric field since its electrical tunability is very low \cite{Rahman2007}. Multi-donor qubits on the other hand, change their total hyperfine significantly when the separations between donors are varied. The total hyperfine value of a 2P dot typically reaches few hundreds of MHz for very close donors but quickly falls below the 1P value beyond $\backsim 3$ nm \cite{Wang2016}. For a qualitative comparison we can take the example of 1P and 2P of the same total hyperfine, i.e. $A=(A_1+A_2)=117$ MHz. The 1P case will be characterized just by the nuclear $\Uparrow\rightarrow\Downarrow$ transition probability $\propto A^2/B^2$. However, the measurement backaction in a 2P dot is dependent on its initial state. For the $\ket{\Uparrow\Uparrow}$ initial state, the effect includes the sum of the flip rates of both donors $\propto (A_1^2+A_2^2)/B^2$. However, regardless of the exact $A_1$ and $A_2$ values, it is still smaller (by $\propto 2A_1A_2/B^2$) than the 1P flip rate. For the $\ket{\Uparrow\Downarrow}$ initial state, we need to account for both flip and flip-flop effects, which would add up to approximately $\propto A_1^2/B^2 + (A_1A_2)^2/(A_1-A_2)^2B^2$. We show the two effects as a function of Stark shift ($A_1-A_2$) in Fig.~\ref{2P_3P} (b) -- numerical and analytical results with dots and dashed lines, respectively. We can see that the nuclear spin flip-flop transitions dominate for small Stark shift, but falls below the single nuclear spin flip rate for $A_1>2A_2$. With the shaded region we show the region where the sum of both flip and flip-flop rates is lower than 1P flip-rate. We can see that this regime starts at $A_1 - A_2 \backsim 35$ MHz, achievable in current devices using electric potentials from the gates \cite{madzik2021precision}. For 2P molecules of smaller total hyperfine, even smaller Stark shifts would be required. This example demonstrates that multi-donor qubits can be more resistant than 1P to measurement backaction effects, despite the presence of additional flip-flop mechanism. As mentioned before, 3P configurations naturally have asymmetry in the hyperfine values due to the donor arrangement so these systems can also be less sensitive to measurement backaction.

 

\section{Conclusion}	
We have studied the effects of measurement backaction in single and multi-donor nuclear spin qubits in Si. We implemented a master equation-based approach that treats the electron tunneling events during different qubit-control pulses as appropriate Lindblad operators and simulate the time evolution. We show that for both single and multi-donor nuclear spin qubits, measurement backaction causes nuclear spin flip whose probability varies as $A^2/B^2$, where, $A$ is the hyperfine constant of the corresponding donor and B is magnetic field. For a particular nuclear spin this transition probability is independent of the hyperfine constants of other nuclear spins present in the multi-donor dot. For these multi-donor qubits, the measurement backaction can also cause flip-flop ($\Uparrow\Downarrow\rightarrow\Downarrow\Uparrow$) transitions between the nuclear spins. This is because the electron also indirectly couples the two nuclear spins that are individually hyperfine-coupled with the electron. This flip-flop probability is a few orders of magnitude larger than the $\Uparrow \rightarrow \Downarrow$ transition probability for small Stark shifts in a 2P donor qubit and becomes smaller for $A_1 > 2A_2$. We show that by positioning donors a few nm apart and operating the qubits at high Stark shift (few tens of MHz), we can minimize the measurement backaction effect on the multi-donor qubit lifetime, to below that of a single donor. These results highlight the ability to engineer multi-donor qubit systems. 

\section*{Acknowledgements}
This research is conducted by the Australian Research Council Centre of Excellence for Quantum Computation and Communication Technology (CE170100012), the US Army Research Office under Contract No. W911NF-17-1-0202 and Silicon Quantum Computing Pty Ltd.

The research is undertaken with the assistance of resources and services from the National Computational Infrastructure (NCI) under an NCMAS 2020 allocation, supported by the Australian Government, and of the computational cluster Katana supported by Research Technology Services at UNSW Sydney.
	
\bibliography{measurement_biblio}

\end{document}


\title{Impact of measurement backaction on nuclear spin qubits in silicon: Supplementary Material}
	
	
	\author{S. Monir*}
	\affiliation{School of Physics, The University of New South Wales, Sydney, NSW 2052, Australia}
	\affiliation{Silicon Quantum Computing Pty Ltd., Level 2, Newton Building, UNSW Sydney, Kensington, NSW 2052, Australia}
	
	\author{E. N. Osika}
	\affiliation{School of Physics, The University of New South Wales, Sydney, NSW 2052, Australia}
	\affiliation{Silicon Quantum Computing Pty Ltd., Level 2, Newton Building, UNSW Sydney, Kensington, NSW 2052, Australia}
	
	\author{S. K. Gorman}
	\affiliation{Silicon Quantum Computing Pty Ltd., Level 2, Newton Building, UNSW Sydney, Kensington, NSW 2052, Australia}
	\affiliation{Centre for Quantum Computation and Communication Technology, School of Physics, University of New South Wales, Sydney, NSW 2052, Australia}
	
	\author{I. Thorvaldson}
	\affiliation{Silicon Quantum Computing Pty Ltd., Level 2, Newton Building, UNSW Sydney, Kensington, NSW 2052, Australia}
	\affiliation{Centre for Quantum Computation and Communication Technology, School of Physics, University of New South Wales, Sydney, NSW 2052, Australia}

	\author{Y.-L. Hsueh}
	\affiliation{School of Physics, The University of New South Wales, Sydney, NSW 2052, Australia}
	\affiliation{Silicon Quantum Computing Pty Ltd., Level 2, Newton Building, UNSW Sydney, Kensington, NSW 2052, Australia}
	
	\author{P. Macha}
	\affiliation{Silicon Quantum Computing Pty Ltd., Level 2, Newton Building, UNSW Sydney, Kensington, NSW 2052, Australia}
	\affiliation{Centre for Quantum Computation and Communication Technology, School of Physics, University of New South Wales, Sydney, NSW 2052, Australia}
	
	\author{L. Kranz}
	\affiliation{Silicon Quantum Computing Pty Ltd., Level 2, Newton Building, UNSW Sydney, Kensington, NSW 2052, Australia}
	\affiliation{Centre for Quantum Computation and Communication Technology, School of Physics, University of New South Wales, Sydney, NSW 2052, Australia}
	
	\author{J. Reiner}
	\affiliation{Silicon Quantum Computing Pty Ltd., Level 2, Newton Building, UNSW Sydney, Kensington, NSW 2052, Australia}
	\affiliation{Centre for Quantum Computation and Communication Technology, School of Physics, University of New South Wales, Sydney, NSW 2052, Australia}
	
	\author{M.Y. Simmons}
	\affiliation{Silicon Quantum Computing Pty Ltd., Level 2, Newton Building, UNSW Sydney, Kensington, NSW 2052, Australia}
	\affiliation{Centre for Quantum Computation and Communication Technology, School of Physics, University of New South Wales, Sydney, NSW 2052, Australia}
	
	\author{R. Rahman*}
	\affiliation{School of Physics, The University of New South Wales, Sydney, NSW 2052, Australia}
	\affiliation{Silicon Quantum Computing Pty Ltd., Level 2, Newton Building, UNSW Sydney, Kensington, NSW 2052, Australia}

\maketitle
\beginsupplement

\onecolumngrid
\subsection{I. Details of the simulation of single-shot readout pulse}
In this supplementary section we discuss the details of the single-shot readout pulse simulations using master equation. We consider a single donor quantum dot (1P1e) whose energy levels are schematically shown in Fig.~\ref{supp_read_res}(a). In order to perform readout, the SET energy level is placed between electron $\ket{\uparrow}$ and $\ket{\downarrow}$ states allowing the electron to tunnel to the SET from the donor dot and then a $\ket{\downarrow}$ electron tunneling back to the donor dot from the SET. The Hamiltonians of the system with the electron on the dot ($H_{donor}$) and on the SET ($H_{zn}$) can be written as,
\begin{align}
    H_{donor} & = \gamma_n \mathbf{I} \cdot \mathbf{B} + \gamma_e \mathbf{S} \cdot \mathbf{B} + A \mathbf{I} \cdot \mathbf{S} \\
    H_{zn} & = \gamma_n \mathbf{I} \cdot \mathbf{B}
\end{align}
where, $\gamma_n (\gamma_e)$ is the nuclear (electron) gyromagnetic ratio, $\mathbf{I}$ ($\mathbf{S}$) is the nuclear (electron) spin operator and $A$ is the Fermi contact hypefine constant. We write the electron as a three-level system ($\{ \ket{\uparrow}, \ket{\downarrow}, \ket{SET} \}$) which combines the two Hamiltonians in the basis $\mathcal{B} = \{\ket{\Uparrow\uparrow}, \ket{\Uparrow\downarrow}, \ket{\Uparrow SET}, \ket{\Downarrow\uparrow}, \ket{\Downarrow\downarrow}, \ket{\Downarrow SET}\}$ as,
\begin{equation*}
	H = \left(\begin{smallmatrix}
	\frac{\omega_e}{2}+\frac{\omega_n}{2}+\frac{A}{4} & 0 & 0 & 0 & 0  & 0 \\
	0 & -\frac{\omega_e}{2}+\frac{\omega_n}{2}-\frac{A}{4} & 0 & \frac{A}{2} & 0 & 0 \\
	0  &  0 & \frac{\omega_n}{2} & 0 & 0 & 0 \\
	0 & \frac{A}{2} & 0  & \frac{\omega_e}{2}-\frac{\omega_n}{2}-\frac{A}{4} & 0 & 0 \\
	0 & 0 & 0 & 0 & -\frac{\omega_e}{2}-\frac{\omega_n}{2}+\frac{A}{4} & 0 \\
	0 & 0 & 0 & 0 & 0 & -\frac{\omega_n}{2} \\
	\end{smallmatrix}\right )
\end{equation*}
where we consider an external magnetic field $\mathbf{B} = (0, 0, B_z)$ and replaced $\gamma_n B_z$($\gamma_e B_z$) with the Larmor frequency $\omega_n$($\omega_e$).

As we can see, the SET and donor parts of the Hamiltonian are independent, thus the eigenstates of the system can be described by two SET states $\ket{\Uparrow SET}$ and $\ket{\Downarrow SET}$ and four donor states which we label $\ket{\mathbf{e}_1} - \ket{\mathbf{e}_4}$ (see Fig.~\ref{supp_read_res}(a)). For the simulations of time evolution, we switch to the eigenbasis and construct the Lindblad operators ($L_\mu$) in that basis. The number of $L_\mu$'s can vary depending on the number of possible ways for the electron to tunnel during a particular spin-control pulse. Fig.~\ref{supp_read_res}(a) shows how the electron tunnels between the  eigenstates of $H_{donor}$ and $H_{zn}$ during the readout. The tunneling itself is a random event but its statistics are characterized by the tunneling time. For the readout pulse in Fig.~\ref{supp_read_res}(a), we have defined the tunneling time $\tau_{\uparrow \rightarrow SET}$ for the electron tunneling to the SET from the qubit and $\tau_{SET \rightarrow \downarrow}$ for electron tunneling to the qubit from the SET. The possible tunneling events in this process can be represented by the following Lindblad operators,
\begin{align}
    L_1 & = \frac{1}{\sqrt{\tau_{\uparrow \rightarrow SET}}} \Bigl( \ket{\Uparrow} \otimes \ket{SET} \Bigr) \bra{\mathbf{e}_4}\mathbb{P}_{donor}^\dagger \\
    L_2 & = \frac{|\braket{\Uparrow \downarrow|\mathbf{e}_3}|}{\sqrt{\tau_{\uparrow \rightarrow SET}}} \Bigl( \ket{\Uparrow} \otimes \ket{SET} \Bigr) \bra{\mathbf{e}_3}\mathbb{P}_{donor}^\dagger \\
    L_3 & = \frac{|\braket{\Downarrow \uparrow | \mathbf{e}_3}|}{\sqrt{\tau_{\uparrow \rightarrow SET}}} \Bigl( \ket{\Downarrow} \otimes \ket{SET} \Bigr) \bra{\mathbf{e}_3}\mathbb{P}_{donor}^\dagger \\
    L_4 & = \frac{1}{\sqrt{\tau_{SET \rightarrow \downarrow}}} \mathbb{I}_{2 \times 2} \otimes \Bigl(\ket{\downarrow}\bra{SET}\Bigr)
\end{align}
Here, the operators $L_1-L_3$ correspond to the electron tunneling out from the qubit to the SET. Operator $L_4$ corresponds to a $\ket{\downarrow}$ electron tunneling in from the SET to the qubit. We assume the tunneling is instantaneous, therefore, the operators preserve the probabilities of nuclear spin states $\ket{\Uparrow}$ and $\ket{\Downarrow}$. The operator $\mathbb{P}_{donor}$ projects the eigenvectors $\ket{\mathbf{e}_1}-\ket{\mathbf{e}_4}$ from the Hilbert space of $H_{donor}$ to the Hilbert space of the combined Hamiltonian $H$ which is spanned by the basis $\mathcal{B}$.

For our simulations we consider a readout pulse with $\tau_{\uparrow \rightarrow SET} = 80$  $\mathrm{\mu s}$, $\tau_{SET \rightarrow \downarrow} = 120$ $\mathrm{\mu s}$ (experimentally realistic values \cite{plathesis}) with a magnetic field $B_z = 1.4$ T. The length of the readout pulse is such that $\tau_{\uparrow \rightarrow SET}+\tau_{SET \rightarrow \downarrow}$ fits well within the simulation window ($1$ ms). With this setup we now simulate the time evolution of the density operator using the master equation in Lindblad form.
\begin{equation} \label{master_eqn}
	    \partial_t \rho = -i[H, \rho]_{-}+\sum_{\mu} \left( L_{\mu} \rho L_{\mu}^{\dagger}-\frac{1}{2}[L_{\mu}^{\dagger}L_{\mu}, \rho]_{+}\right)
\end{equation}
The density operator $\rho$ is of the combined electron-nuclear system. In order to calculate any specific spin state probability (e.g., nuclear $\ket{\Uparrow}$ state) we take expectation values of appropriate projection operators.
    
\subsection{II. Benchmarking with previous experiment}
\begin{figure*}
	\centering
	\includegraphics[width = 15 cm]{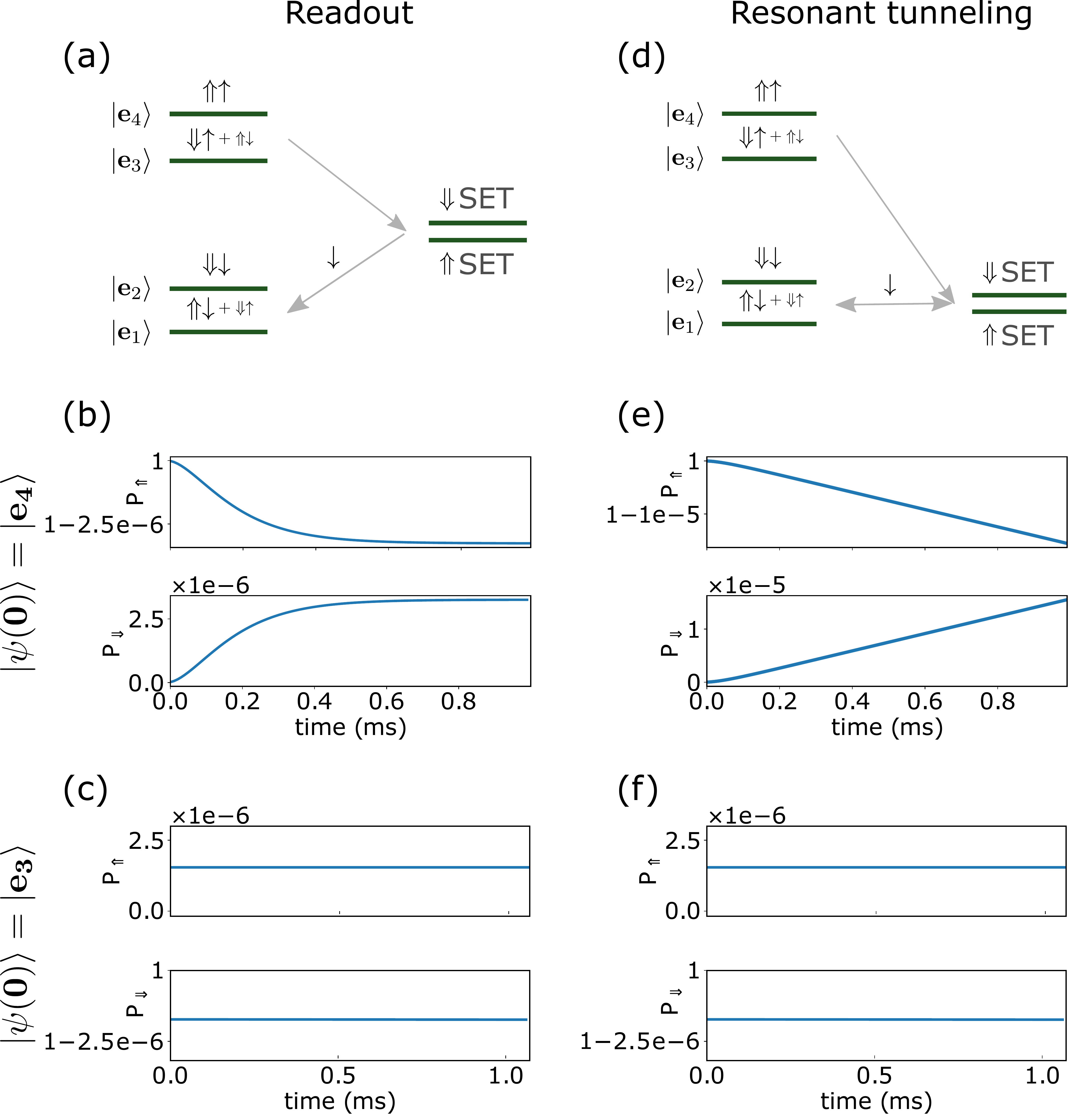}%
	\caption{\textbf{Nuclear spin flip probabilities during the readout and resonant tunneling pulses}. a) Electron tunneling in a 1P1e qubit during the readout pulse. The labels $\ket{\mathbf{e}_1} - \ket{\mathbf{e}_4}$ represent eigenstates of the 1P1e system. The arrows show the nuclear-electron spin states forming the eigenstates, with the smaller arrows (i.e $\Downarrow\uparrow$ in $\ket{\mathbf{e}_1}$ and $\Uparrow\downarrow$ in $\ket{\mathbf{e}_3}$) representing typically very small admixtures  of nuclear-electron spin configurations introduced by hyperfine interaction. Probabilities of nuclear spin states as a function of time during readout when the system is initialized in b) $\ket{\mathbf{e}_4}$ and c) $\ket{\mathbf{e}_3}$. d) Electron tunneling during the resonant tunneling phase. e) and f) are nuclear spin probabilities during resonant tunneling phase for initial states same as in b) and c) respectively. 
    \label{supp_read_res}}
\end{figure*}

We have verified the method presented in this paper by reproducing the single-donor experimental data from Ref.~\cite{plathesis}. In the experiment the resonant tunneling pulses were used to study the effect of measurement backaction. We now discuss the resonant tunneling pulse in Fig.~\ref{supp_read_res}(d) and the difference in modelling between the readout and resonant tunneling pulses in the following.

The resonant tunneling pulse was introduced in the ESR experiment in order to study the effects of measurement backaction in Ref.~\cite{plathesis}. During the resonant tunneling pulse the SET chemical potential is set equal to the majority spin $\ket{\downarrow}$ electronic level, i.e. $\backsim \ket{\mathbf{e}_1}, \ket{\mathbf{e}_2}$, so the $\ket{\downarrow}$ electron tunnels back and forth between the SET and the qubit. Fig.~\ref{supp_read_res}(d) shows the energy levels of a 1P1e qubit during a resonant tunneling pulse. Since the SET and $\ket{\downarrow}$ levels are in resonance with each other, the electron tunnels between the two multiple times. Fig.~\ref{supp_read_res}(e,f) shows the simulation results of the resonant tunneling pulse, using the same $\tau$ and $B_z$ parameters as for the readout, but with an additional path available for the electron in the majority $\ket{\downarrow}$ levels tunneling from the dot to the SET with $\tau_{\downarrow \rightarrow SET} = 80$ $\mathrm{\mu s}$.

We show how the nuclear spins are affected by measurement backaction by plotting $P_\Uparrow$ and $P_\Downarrow$ probabilities both for readout and resonant tunneling pulses - see Fig.~\ref{supp_read_res}(b, c) and (e, f), respectively. Fig.~\ref{supp_read_res}(b,e) and (c,f) correspond to the simulations started with initial states of $\ket{\mathbf{e}_4}$ and $\ket{\mathbf{e}_3}$, respectively. If we start the readout pulse with the $\ket{\mathbf{e}_4}$ state, the first tunneling event where the electron goes to the SET leaves the system in its eigenstate. However, the $\ket{\downarrow}$ electron tunneling back to the qubit mixes the $\ket{\Uparrow}$ and $\ket{\Downarrow}$ nuclear spins through the hyperfine interaction which introduces a small probability of $\Uparrow\rightarrow\Downarrow$ transition. If the initial state is $\ket{\mathbf{e}_3}$, the small admixture of $\ket{\Uparrow\downarrow}$ gives a small nuclear $\ket{\Uparrow}$ state probability $P_\Uparrow$. However, this probability stays the same throughout the readout pulse (see Fig.~\ref{supp_read_res}(c)). In addition to this, as we mentioned in the main text, the $\Downarrow \rightarrow \Uparrow$ transition is dominated by the hyperfine mediated relaxation mechanism, therefore, the effect of the measurement backaction does not give any signature in the $\Downarrow \rightarrow \Uparrow$ transition \cite{plathesis}.

In the case of the resonant tunneling pulse, each tunneling of the $\ket{\downarrow}$ electron to the qubit introduces a mixture of $\ket{\Uparrow\downarrow}$ and $\ket{\Downarrow\uparrow}$ states through the hyperfine interaction, resulting in a finite nuclear spin $\Uparrow\rightarrow\Downarrow$ transition probability. These probabilities add up each time, as is seen from Fig.~\ref{supp_read_res}(e). With $\tau_{\uparrow \rightarrow SET} + \tau_{SET \rightarrow \downarrow} = 200$ $\mathrm{\mu s}$, the electron tunnels in and out of the dot about 5 times within a duration of 1~ms. Hence the $\Uparrow \rightarrow \Downarrow$ transition rate becomes $\backsim 10^{-5}$, 5 times as large as in the case of the readout pulse. During this pulse, the nuclear spin transition becomes a one-way process ($\Uparrow \rightarrow \Downarrow$).

\begin{figure*}[t]
\centering
\includegraphics[width = 8 cm]{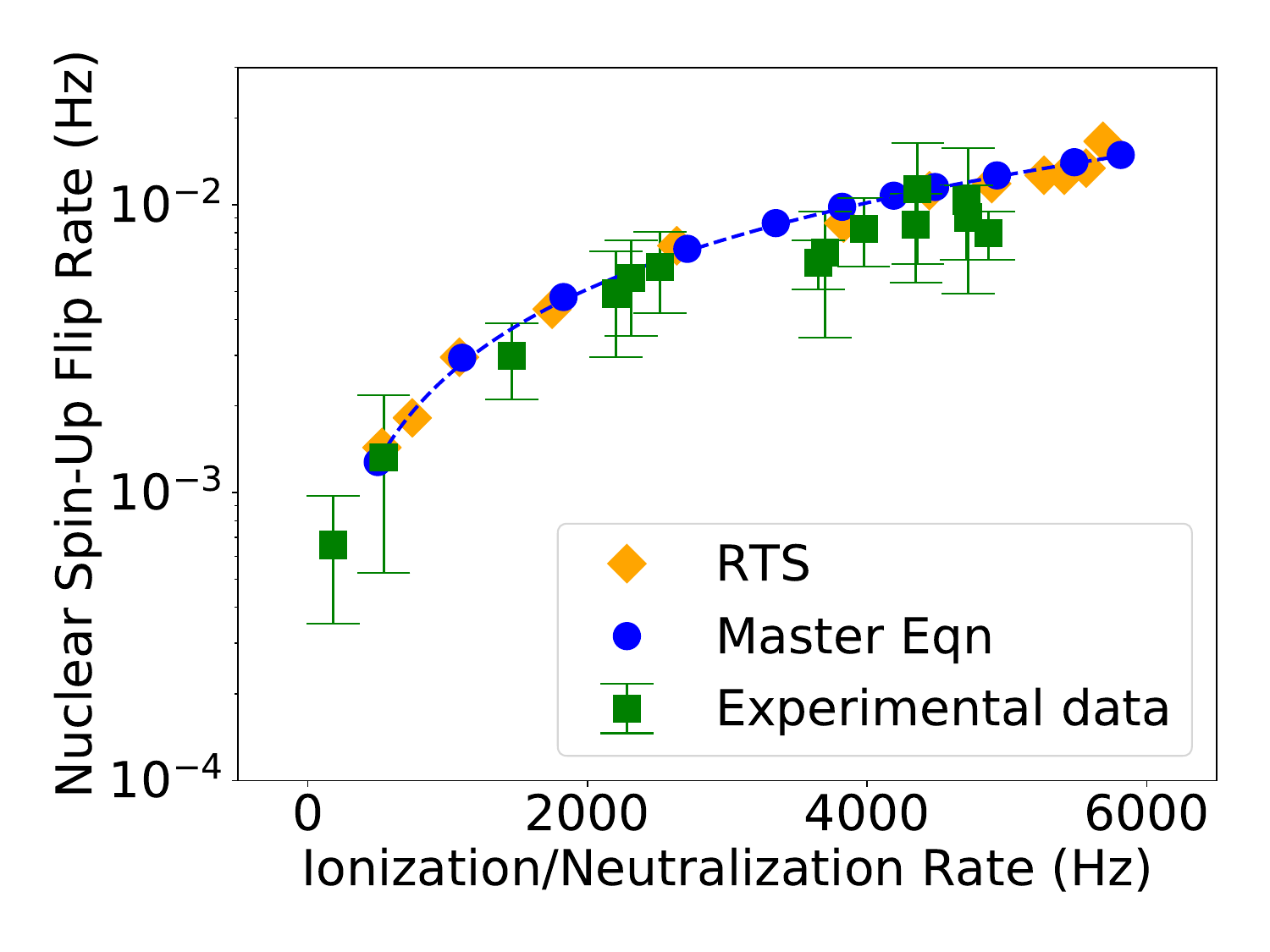}%
\caption{\textbf{Benchmarking our model with previous results}. The plot shows nuclear spin $\Uparrow\rightarrow\Downarrow$ flip rates as a function of ionization/neutralization rate as calculated using master equation approach presented in this paper (blue points) and as presented in Ref.~\cite{plathesis} -- measured in experiment (green squares) and simulated with RTS (orange diamonds).
\label{pla_exp}}
\end{figure*}

We can see in Fig.~\ref{pla_exp} that our results agree very well both with experimental and RTS results from Ref.~\cite{plathesis}. Additionally, in contrast to RTS simulations, the master-equation method calculates the theoretical average of an infinite number of experimental runs in a one-pass simulation, allowing faster calculations with more accuracy. In the main text we focus on the readout process only, as the resonant tunneling pulses are not typically included in qubit control protocols.

The point here to note is that the measurement backaction is mainly driven by contact term of the hyperfine interaction. Even when neglecting the anisotropic terms we reproduce the experimental results (see Fig.~\ref{pla_exp}) very well. The anisotropic hyperfine terms are generally three orders of magnitude smaller than the contact hyperfine term \cite{rahman2009}, therefore, even in multi-donor quantum dots where the wavefunctions are highly anisotropic, the measurement backaction is dominated by the contact hyperfine term.

\subsection{III. Dipole-dipole interaction}
	In addition to electron mediated effects, the nuclear spins are also directly coupled through the dipole-dipole interaction. This mechanism can also cause flip-flop transitions between the nuclear spins when the electron is not present on the qubit. In the presence of the electron, nuclear spins are strongly polarized by the hyperfine interaction and the transitions due to the nuclear spin dipole-dipole interaction are suppressed. The strength of the nuclear dipole-dipole interaction depends on the inter-donor separation with $1/r^3$ with $r$ being the distance between the donors along the separation axis. Above a separation of 3 lattice constants the dipole-dipole strength falls below 5Hz. Typical donor separations observed in the experiment result in total hyperfine interactions of about 200-600 MHz, and for the same separations the donor nuclear spin dipole-dipole interaction can vary in the range of a few Hz to a few tens of Hz. Because this interaction is only active during the period between the electron having tunneled out and a new electron tunneling in - which is typically from a few $\mu s$ \cite{keith2020fast} to 1 ms \cite{Hile2018} - the overall spin-flip-flop probability is between $10^{-6}$ to $10^{-3}$. We can see from Fig. 3(a) of the main text that for large Stark shifts of around tens of MHz, the flip-flop rate due to measurement backaction and nuclear dipole-dipole interaction are comparable. However, for smaller Stark shift the flip-flop process induced by measurement backaction can dominate over the dipole-dipole induced one. 
	
	We showed in the Letter that in a 2P1e system with a total hyperfine value of $\backsim$ 117 MHz, the measurement backaction can be made smaller than that of a 1P1e system. However, the nuclear spin dipole-dipole related flip-flop process would now add to the effects, potentially hindering the lifetime of the 2P system. To ensure that the sum of all flip and flip-flop rates does not surpass the single donor value, we can minimize the time when the donors are ionized. The donor separation for 117 MHz total hyperfine is $\backsim$ 3-4 nm \cite{Wang2016}. At this distance the nuclear spin dipole-dipole interaction is of the order of 0.1 - 1 Hz. To ensure that the nuclear spin transition probabilities in multi-donor dots remain below that of a single donor, we would require that the time when the multi-donor dot has no electron is kept in the range 10-100 $\mu$s.

\bibliography{supp_bib}